\documentclass[prd,reprint,no title page,twoside,onecolumn,showpacs,
               nofootinbib,longbibliography,amsmath,amssymb,
               floatfix]{revtex4-1}

\usepackage{graphicx}   
\usepackage{bm}         
\usepackage{mathtools}
\usepackage{bigints}

\mathchardef\ordinarycolon\mathcode`\:                     %
\def\vcentcolon{\mathrel{\mathop\ordinarycolon}}           
\providecommand*\coloneqq{{\;\vcentcolon\mkern-0.8mu=\;}}  
\DeclareMathOperator{\sgn}{sgn}

\begin{document}

\enlargethispage{1.0in}

\title{Geodesic Particle Paths Inside a Nonrotating, Homogeneous, Spherical
Body}

\author{Homer G.~Ellis}
\affiliation{Department of Mathematics, University of Colorado at Boulder,
Boulder, Colorado 80309}

\date{October 16, 2012}

\begin{abstract}
Proceeding from a solution of field equations that are improved versions of
Einstein's nonvacuum gravitational field equations one is able to calculate
precisely the trajectories of particles traveling inside a nonrotating,
homogeneous, spherical body.  Application of the results to the conditions of
recent measurements of neutrino flight times between a source point A at CERN's European Laboratory for Particle Physics and a point B in either of two
detectors (ICARUS or OPERA) at LNGS (Laboratori Nazionale del Gran Sasso),
separated by a euclidean distance $d(\text{A},\text{B}) = 731$ km, predicts for
the flight time $T_\nu$ from A to B of a 2 eV neutrino launched with energy
17 GeV that, as measured by a clock at B synchronized to a similar clock at A,
$T_\nu \approx d/c + 9.3 \times 10^{-16}$ sec.  But as measured by inertial
observers along the path the flight time
$\bar T_\nu \approx d/c - 2.6 \times 10^{-9}$ sec and the path length
$L_\nu \approx d - 8.4 \times 10^{-7}$ m, which yields
$L_\nu/\bar T_\nu \approx c + 321$ m/sec for the average inertially referenced
speed of the neutrino from A to B.
\end{abstract}


\maketitle

\section{THE INTERIOR METRIC OF THE SPHERICAL BODY}

In a previous paper I derived a space-time metric for the gravitational field
inside a nonrotating, homogeneous, spherical ball $\cal B$, matched as smoothly
as possible at the surface of $\cal B$ to a Schwarzschild exterior
metric~\cite{elli1}.  This new metric is a solution of the field equations
\begin{equation}
{\bm R}_{\alpha \beta} - {\textstyle\frac12} {\bm R} \, g_{\alpha \beta}
  = -\frac{4 \pi \kappa}{c^2} \mu \, g_{\alpha \beta} \, ,
\label{eqn1}
\end{equation}
which come from the variational principle
\begin{equation}
\delta \! \bigintssss \!\! \left({\bm R}
 - \frac{8 \pi \kappa}{c^2} \mu \right) |g|^{\frac12} \, d^4\!x = 0 \, ,
\label{eqn2}
\end{equation}
in which $\kappa$ is Newton's gravitational constant and $\mu$ is {\it active}
gravitational mass density, and which is the most natural extension to the
general relativity setting of the variational principle 
$\delta \! \int (|\nabla V|^2 + 8 \pi \kappa \mu V) \, d^3\!x = 0$ that
produces the Poisson equation $\nabla^2 V = 4 \pi \kappa \mu$ for the newtonian
gravitational potential~$V$.\footnote{Justification for the complete version of
this variational principle and the field equations that it implies can be seen
in~\cite{elli2} and, in greater detail, in~\cite{elli3}.}  Because these field
equations differ from those that Einstein postulated in
1916~(\cite{eins1},\S 16,\S 19) (and that have been taken as gospel ever
since), the interior solution they yielded in~\cite{elli1} differs
significantly from the Schwarzschild interior solution~\cite{schw1,elli1}.  In
particular, the metric can be expressed entirely in terms of rational functions
of the radial coordinate, which makes feasible a relatively straightforward
analysis of its geodesics and thereby allows computations of flight times and
travel distances of test particles such as photons and neutrinos following
geodesics between points on the surface of the spherical ball considered as
representing Earth (provided with tunnels for the photons to travel in affected
only by gravity).

The metric takes the proper-time forms
\begin{align}
d\tau^2
 &= [1 - f^2(\rho)] \, dt^2 -
     \frac{1}{c^2} \, [1 - f^2(\rho)]^{-1} \, d\rho^2 -
     \frac{1}{c^2} \, r^2(\rho) \, d\Omega^2
\label{eqn3} \\
 &= d{\bar t}\,^2 - \frac{1}{c^2} \, \left[d\rho
                          - f(\rho) \, c \, d{\bar t} \, \right]^2
                  - \frac{1}{c^2} \, r^2 (\rho) \, d\Omega^2 \, ,
\label{eqn4}
\end{align} 
in which $\displaystyle \bar t \coloneqq t
          - (1/c) \! \int \! f(\rho)[1 - f^2(\rho)]^{-1} \, d\rho$,
$r(\rho) = \lambda (\rho - \rho_0)$ (the areal radius of a sphere of
geodesic radius $\rho - \rho_0$ in a generic $\bar t$ time slice, normalized
so that $r(\rho) = R$ when $\rho = R$, the radius of the ball $\cal B$, which
makes $\rho_0 = (1 - 1/\lambda) \, R$),
\begin{equation}
1 - f^2
 = \frac{1}{\lambda^2}
   \left(1 + \frac{\lambda \kappa M}{c^2 R} \,\frac{r^2}{R^2}\right)
 = \frac{1}{\lambda^2}
   \left(1 + \frac{\lambda m}{R^3} \, r^2\right) \, ,
\label{eqn5}
\end{equation}
$f \coloneqq -\sqrt{f^2}$, $M$ is the active gravitational mass of $\cal B$,
$m = \kappa M/c^2$ ($= M$ in geometric units), and the dimensionless parameter
\begin{equation}
\lambda = \frac{m + \sqrt{m^2 + 4 R (R - 2 m)}}{2 (R - 2 m)} \, .
\label{eqn6}
\end{equation}
It is assumed that $R > 2 m$ (the Schwarzschild radius of $\cal B$), from which
it follows that $\lambda > 1$ and $\rho_0 >0$.

The vector field
$\partial_{\,\bar t} + f(\rho) \, c \, \partial_\rho$ is the velocity field of
a cloud of inertial observers free-falling from rest at $\rho = \infty$;
the time $\bar t$ runs at the same rate as their proper times.  The geometry of
space as seen by these observers is described by the metric of a time-slice
$\Sigma_{\bar t}$ of constant $\bar t$, namely
$d\sigma^2 = d\rho^2 + r^2(\rho) d\Omega^2$.  Rather than being flat euclidean,
as in a corresponding slice of the Schwarzschild exterior metric, in which
$r(\rho) = \rho$ (the geodesic distance from the point singularity where
$r$ would be 0 if the ball $\cal B$ were collapsed to that point), they are
`hyperconical' in that $r(\rho) = \lambda (\rho - \rho_0) > \rho - \rho_0$ (the
geodesic distance from the center of $\cal B$ where $\rho = \rho_0$ and $r$ is
in fact 0 --- the vertex of the `hypercone').  Although $\Sigma_{\bar t}$ has a
curvature singularity at the vertex of the hypercone (the $\vartheta\varphi$
sectional curvature being $(1 - r'^{\,2})/r^2 = -(\lambda^2 - 1)/r^2$), the
full space-time manifold does not, as one can show that inside $\cal B$ every
one of the sectional curvatures is the same $-\lambda m/R^3$.

\section{GEODESICS INSIDE THE SPHERICAL BODY $\cal B$}

To study geodesics inside the spherical ball $\cal B$, in particular geodesics
between two points A and B on the surface of $\cal B$, let us orient the
spherical polar coordinate system of Eqs.~(\ref{eqn3}) and (\ref{eqn4}) so that
A and B are on a longitude and equidistant from the equator (A to the north and
B to the south), and in place of the usual colatitude coordinate $\vartheta$
use the latitude coordinate $\theta$, the two related by
$\theta = \pi/2 - \vartheta$.
Then
$d\Omega^2 = d\vartheta^2 + (\sin \vartheta)^2 d\varphi^2
           = d\theta^2 + (\cos\theta)^2 d\varphi^2$, and
$\theta = \delta$ at A and $-\delta$ at B, where
$\delta = \sin^{-1}(d/2 R)$ and $d$ is the euclidean distance from A to B.

For every affinely parametrized geodesic path in $\cal B$ with longitude
$\varphi$ fixed there are three additional constants of the motion, namely,
\begin{align}
       h &\coloneqq \frac{1}{c} \, r^2 \dot\vartheta
          = -\frac{1}{c} \, r^2 \dot\theta \, ,
\label{eqn7} \\
       k &\coloneqq (1 - f^2) \, \dot t
          = \dot{\bar t} + \frac{f}{c} (\dot\rho - f c \, \dot{\bar t} \,) \, ,
\label{eqn8}
\vspace{-90pt}
\shortintertext{\noindent and}
\epsilon &\coloneqq (1 - f^2) \, {\dot t}^2
                    - \frac{1}{c^2} \, \frac{1}{1 - f^2} \, \dot\rho^2
                    - \frac{1}{c^2} \, r^2 \dot\theta^2 \, ,
\label{eqn9} \\
         &= \dot{\bar t}\,^2 - \frac{1}{c^2} \, \left(\dot\rho
                                   - f c \, \dot{\bar t} \, \right)^{\!2}
                           - \frac{1}{c^2} \, r^2 \dot\theta^2 \, ,
\label{eqn10}
\end{align}
where $\epsilon = 1, 0, -1$ according as the path is timelike (parametrized by
arclength), lightlike, or spacelike (parametrized by arclength).  From
these equations and Eq.~(\ref{eqn5}) follow
\begin{align}
\dot\rho^2
 &= c^2 \left[k^2 - (1 - f^2)\left(\frac{h^2}{r^2} + \epsilon\right)\right]
\label{eqn11} \\
 &= -\frac{c^2}{r^2}
     \left[\epsilon \, \frac{m}{\lambda R^3} \, r^4
           - \frac{(\lambda^2 k^2 - \epsilon) R^3 - \lambda m h^2}
                  {\lambda^2 R^3} \, r^2
           + \frac{h^2}{\lambda^2}\right] \, .
\label{eqn12}
\end{align}

\subsection{Lightlike geodesics}

For a lightlike geodesic $\epsilon = 0$ and Eq.~(\ref{eqn12}) reduces to
\begin{equation}
\dot\rho^2
 = \frac{c^2}{r^2}
   \left[\frac{\lambda^2 k^2 R^3 - \lambda m h^2}{\lambda^2 R^3} \, r^2
         - \frac{h^2}{\lambda^2}\right] \, .
\label{eqn13} 
\end{equation}
If $\lambda^2 k^2 R^3 - \lambda m h^2 \leq 0$, then nonnegativity of
$\dot\rho^2$ forces $h = 0$, which then forces $k = 0$; the geodesic is
degenerate, comprising a single event.  If
$\lambda^2 k^2 R^3 - \lambda m h^2 > 0$, then either $h = 0$, in which case
$\dot\theta = 0$ and the geodesic traces out a diameter of $\cal B$, or else
$h \neq 0$ and Eq.~(\ref{eqn13}), which precludes $k = 0$, when combined with
Eq.~(\ref{eqn7}) yields
\begin{equation}
\left(\frac{dr}{d\theta}\right)^{\!2}
 = \left(\frac{dr}{d\rho}\right)^{\!2} \left(\frac{d\rho}{d\theta}\right)^{\!2}
 = \lambda^2 \frac{\dot\rho^2}{\dot\theta^2}
 = \frac{r^2 (r^2 - r_0^2)}{r_0^2} \, ,
\label{eqn14}
\end{equation}
where
\begin{equation}
r_0 = \sqrt{\frac{R^3 (h/k)^2}{\lambda^2 R^3 - \lambda m (h/k)^2}} \, .
\label{eqn15}
\end{equation}
As a lightlike particle travels from A to B (through a tunnel created for its
passage) the value of $r$ decreases from $R$ at A to $r_0$ at the halfway point
H, where $\theta = 0$, then increases back to $R$ at B.  From A to H, when
$\theta$ is positive, $d\theta/dr > 0$, and from H to B, when $\theta$ is
negative, $d\theta/dr < 0$, so
\begin{equation}
\theta
 = \sgn(\theta) \!\! \bigintssss_{r_0}^r \!\! \frac{d\theta}{ds} \, ds
 = \sgn(\theta) \!\! \bigintssss_{r_0}^r \!\! \frac{r_0}
                                                   {s \sqrt{s^2 - r_0^2}} \, ds
 = \sgn(\theta) \sec^{-1} \! \left(\frac{r}{r_0}\right) \, ,
\label{eqn16}
\end{equation}
and therefore $r = r_0 \sec(\sgn(\theta) \, \theta) = r_0 \sec\theta$.  This
describes a trajectory that in euclidean geometry, where $x = r \cos\theta$
and $y = r \sin\theta$, would be a straight line interval from A to B.
To determine $r_0$, note that, at A, $r = R$ and $\theta = \delta$, so that
$R = r_0 \sec(\delta)$, thus
$r_0 = R \cos(\delta) = R \cos(\sin^{-1}(d/2 R)) = \sqrt{R^2 - (d/2)^2}$.  The
geodesic distance from the center $\cal C$ of $\cal B$ to the halfway point H
is then $R \cos(\delta)/\lambda$.

From Eqs.~(\ref{eqn9}), (\ref{eqn14}), and (\ref{eqn5}), together with
$r = r_0 \sec\theta$, it is relatively straightforward to calculate that
\begin{equation}
\left(\frac{dt}{d\theta}\right)^{\!2}
 = \frac{\lambda^2 r_0^2 [1 + (\lambda m/R^3) r_0^2]}
        {c^2 [(\cos\theta)^2 + (\lambda m/R^3) r_0^2]^2} \, . 
\label{eqn17}
\end{equation}
If the particle starts from A at time $t_{\text A}$ and arrives at B at time
$t_{\text B}$, then, because $t$ is increasing as $\theta$ is decreasing,
\begin{align}
t_{\text B} - t_{\text A}
 &= \bigintssss_{\delta}^{-\delta}
     \!\! \frac{dt}{d\theta} \, d\theta
  = \frac{1}{c} \bigintsss_{-\delta}^{\delta} 
                 \!\! \frac{\lambda r_0 \sqrt{1 + (\lambda m/R^3) r_0^2}}
                      {(\cos\theta)^2 + (\lambda m/R^3) r_0^2} \, d\theta
\label{eqn18} \\
 &= \frac{2}{c} \sqrt{\frac{\lambda R^3}{m}}
                \tan^{-1} \! \left(\sqrt{\frac{\lambda m r_0^2}
                                        {R^3 + \lambda m r_0^2}}
                                   \, \tan(\delta)\right)
\label{eqn19} \\
 &= \frac{2}{c} \sqrt{\frac{\lambda R^3}{m}}
                \tan^{-1} \! \left(\sqrt{\frac{\lambda m}
                                        {R + \lambda m \cos^2(\delta)}} 
                                   \, \sin(\delta)\right) \, .
\label{eqn20}
\end{align}
An accurate clock at B perfectly synchronized with a matching clock at A would
record the flight time $T_{\epsilon = 0}$ of the particle as the proper time
elapsed at B since the particle left A, that is to say,
$T_{\epsilon = 0} = \sqrt{1 - f^2(R)} \, (t_{\text B} - t_{\text A})$.

In a similar manner, starting from the metric
$d\sigma^2 = d\rho^2 + r^2(\rho) d\Omega^2$ of $\Sigma_{\bar t}$,
one finds that
\begin{equation}
\left(\frac{d\sigma}{d\theta}\right)^{\!2}
 = \frac{r_0^2 [\lambda^2 + (\tan\theta)^2]}{\lambda^2 (\cos\theta)^2} \, ,
\label{eqn21}
\end{equation}
and then that the length $L_{\epsilon = 0}$ of the path followed by the
particle is given by
\begin{equation}
L_{\epsilon = 0}
= \frac{R \cos(\delta)}{\lambda} \!
   \bigintss_{-\delta}^{\delta} \!\!\!\!\!
    \frac{\sqrt{\lambda^2 + (\tan\theta)^2}}{\cos\theta} \, d\theta \, ,
\label{eqn22}
\end{equation}
which involves elliptic integrals so must be integrated numerically.  It is
straightforward to show analytically that the result will lie between
$d/\lambda$ and $d$.

To find the (average) speed of a photon on a flight from A to B, as measured by
inertial observers free-falling from rest at $\rho = \infty$, we need
$\bar t_{\text B} - \bar t_{\text A}$.  From
$\displaystyle \bar t
 \coloneqq t - (1/c) \! \int \! f(\rho)[1 - f^2(\rho)]^{-1} \, d\rho
         = t - (1/\lambda c) \! \int \! f(1 - f^2)^{-1} \, dr
         = t - (1/\lambda c) \! \int \! f(1 - f^2)^{-1} \, (dr/d\theta)
                                                        \, d\theta$,
(anti)symmetry between the flight from A to H and the flight from H to B,
Eq.~(\ref{eqn5}), and $f \coloneqq -\sqrt{f^2}$, we get that
\begin{align}
\bar T_{\epsilon = 0} \coloneqq \bar t_{\text B} - \bar t_{\text A}
 &= t_{\text B} - t_{\text A}
    - \frac{1}{\lambda c} \! \bigintss_{\delta}^{-\delta} \!\!\!\!\!\!
                             \frac{f}{1 - f^2} \, \frac{dr}{d\theta} \, d\theta
  = t_{\text B} - t_{\text A}
    + \frac{1}{\lambda c} \! \bigintss_{-\delta}^{\delta} \!\!\!\!
                             \frac{f}{1 - f^2} \, \frac{dr}{d\theta} \, d\theta
\label{eqn23} \\
 &= t_{\text B} - t_{\text A}
    + \frac{2}{\lambda c} \! \bigintss_0^{\delta} \!\!\!\!
                             \frac{f}{1 - f^2} \, \frac{dr}{d\theta} \, d\theta
  = t_{\text B} - t_{\text A}
    + \frac{2}{\lambda c} \! \bigintss_{r_0}^R \!\!\!\!
                             \frac{f}{1 - f^2} \, dr
\label{eqn24} \\
 &= t_{\text B} - t_{\text A}
    - \frac{2}{c} \! \bigintss_{r_0}^R \!\!\!
                     \frac{\sqrt{\lambda^2 - 1 - (\lambda m/R^3) r^2}}
                          {1 + (\lambda m/R^3) r^2} \, dr \, . 
\label{eqn25}
\end{align}

\subsection*{Application: Photon flight from CERN to Gran Sasso}

These results can be immediately applied to a situation of recent interest by
choosing for $d$ the euclidean distance between the end points A (CERN) and 
B (Gran Sasso) of the neutrino flight path in the experiments described in
\cite{opera1} and \cite{icarus}, that is, $d = 731$ km (rounded).  With
$M = 5.9722 \times 10^{24}$ kg (the active gravitational mass of Earth) and
$R = 6.3710 \times 10^3$ km (the mean radius of Earth) the outcomes, compared
to $d$, $c$, and $d/c = 0.00244$ sec, are
\begin{itemize}
\item photon's flight time measured by clock at B sychronized with clock at~A:
      \begin{equation}
      T_{\epsilon = 0} = d/c + 9.31085 \times 10^{-16} \text{ sec}
      \label{eqn26}
      \end{equation}
\item photon's flight time from A to B as measured by free-falling inertial
      observers:
      \begin{equation}
      \bar T_{\epsilon = 0} = d/c - 2.61131 \times 10^{-9} \text{ sec}
      \label{eqn27}
      \end{equation}
\item length of photon's flight path from A to B as measured by free-falling
      inertial observers:
      \begin{equation}
      L_{\epsilon = 0} = d - 8.38501 \times 10^{-5} \text{ cm}
                       = d - 0.838501 \; \mu\text{m}
      \label{eqn28}
      \end{equation}
\item average speed of photon in flight from A to B as measured by
      free-falling inertial observers:
      \begin{equation}
      \frac{L_{\epsilon = 0}}{\bar T_{\epsilon = 0}}
      = c + 0.32106 \text{ km/sec}
      \quad (= 1.00000107093 \, c = (1 + 1.07093 \times 10^{-6}) \, c)
      \label{eqn29}
      \end{equation}
\end{itemize}
The corresponding numbers for neutrino flights from A to~B will be found in the
next section.

\subsection{Timelike geodesics}

For a timelike geodesic parametrized by the arclength parameter $\tau$,
$\epsilon = 1$ and Eq.~(\ref{eqn12}) reduces to
\begin{equation}
\dot\rho^2
 = -\frac{c^2 m}{\lambda R^3 r^2}
    \left(r^4 - 2 \, \alpha \, r^2 + \frac{R^3 h^2}{\lambda m}\right) \, ,
\label{eqn30}
\end{equation}
where
\begin{equation}
\alpha = \frac{(\lambda^2 k^2 - 1) R^3- \lambda m h^2}{2 \lambda m} \, .
\label{eqn31}
\end{equation}
If $\alpha \leq 0$, then nonnegativity of $\dot\rho^2$ forces $h = 0$
and $r = 0$, so a particle on this path would be forever stuck at the center
$\cal C$ of $\cal B$.  If $\alpha > 0$, then Eq.~(\ref{eqn30}) becomes
\begin{equation}
\dot\rho^2 = \frac{c^2 m}{\lambda R^3 r^2} (r^2 - a^2)(b^2 - r^2) \, ,
\label{eqn32}
\end{equation}
where $a = \sqrt{\alpha - \beta}$, $b = \sqrt{\alpha + \beta}$, and
$\beta = \sqrt{\alpha^2 - R^3 h^2/\lambda m}$.  If $h = 0$, then
$\dot\theta = 0$ and $a = 0$, and the particle's position oscillates along a
diameter of $\cal B$ between extremes at $r = b$, if $b \leq R$, or else enters
$\cal B$ at one end of a diameter and exits at the other end (in either case,
$r = b \, |\!\sin(\sqrt{\lambda m/R^3} \, c \, \tau)|$).  If $h \neq 0$, then
$a > 0$ and Eq.~(\ref{eqn32}) combined with
Eq.~(\ref{eqn7}) produces
\begin{equation}
\left(\frac{dr}{d\theta}\right)^{\!2}
 = \lambda^2 \frac{\dot\rho^2}{\dot\theta^2}
 = \frac{\lambda m}{R^3 h^2} \, r^2 (r^2 - a^2)(b^2 - r^2)
 = \frac{r^2 (r^2 - a^2)(b^2 - r^2)}{a^2 \, b^2} \, ,
\label{eqn33}
\end{equation}
This implies that
$a \leq r \leq b$, and if $r = a$ when $\theta = 0$, then, as in
Eq.~(\ref{eqn16}),
\begin{align}
\theta
 &= \sgn(\theta) \!\! \bigintssss_a^r \!\!\! \frac{d\theta}{ds} \, ds
  = \sgn(\theta) \!\!\! \bigintss_a^r \!\!\!\!\!
                 \frac{a \, b}{s \sqrt{(b^2 - s^2)(s^2 - a^2)}} \, ds
\label{eqn34} \\
 &= \sgn(\theta)
    \tan^{-1} \!\! \left(\frac{b}{a} \sqrt{\frac{r^2 - a^2}
                                                {b^2 - r^2}}\right) \, ,
\label{eqn35}
\end{align}
from which follows
\begin{equation}
r = \frac{a \, b}
         {\sqrt{a^2 \, (\sin\theta)^2 + b^2 \, (\cos\theta)^2}} \, .
\label{eqn36}
\end{equation}
If $b \leq R$, this describes an oval orbit within $\cal B$ that in euclidean
geometry, where $x = r \cos\theta$ and $y = r \sin\theta$, would be an ellipse
centered on $\cal C$, with minor ($x$) axis of length $2a$ and major ($y$) axis
of length $2b$.  If $b > R$, the trajectory described is an arc of such an oval
connecting points A and B on the surface of $\cal B$ at which $r = R$ and
$\theta = \pm \delta = \pm\sin^{-1}(d/2 R)$, where $d$ is the euclidean
distance from A to B.  Used in Eq.~(\ref{eqn36}), $r = R$ and $\theta = \delta$
produce
\begin{equation}
a \, b
 = R {\sqrt{a^2 \, (\sin\delta)^2 +
            b^2 \, (\cos\delta)^2}}
\label{eqn37}
\end{equation}
as an initial condition to help fix the constants $h$ and $k$.  A useful
consequence of this equation and
$a^2 b^2 = \alpha^2 - \beta^2 = R^3 h^2/\lambda m$ is that
\begin{equation}
h^2 = \frac{\lambda m}{R^3} a^4 \frac{R^2 - r_0^2}{a^2 - r_0^2} \, ,
\label{eqn38}
\end{equation}
from which follows that $a > r_0$ and $h \to \infty$ as $a \to r_0$ (and vice
versa).

From $d\theta/d\tau = \dot\theta = -c \, h/r^2$ and Eq.~(\ref{eqn36}) one can
find the proper time elapsed on the particle's clock as follows:
\begin{align}
\tau_{\text B} - \tau_{\text A}
  = \int_{\tau_{\text A}}^{\tau_{\text B}} \!\!\! d\tau
  = -\frac{1}{c \, h} \int_{\delta}^{-\delta} \!\!\!\! r^2 \, d\theta
 &= \frac{a^2 b^2}{c \, h}
     \bigintssss_{-\delta}^{\delta} \!
      \frac{1}{a^2 \, (\sin\theta)^2 + b^2 \, (\cos\theta)^2} \, d\theta
\label{eqn39} \\
 &= \frac{2 \, a b}{c \, h}
    \tan^{-1} \! \left(\frac{a}{b} \tan(\delta)\right) \, .
\label{eqn40}
\end{align}
To find the proper time elapsed on a clock at B synchronized with a clock at A
requires computing $t_{\text B} - t_{\text A}$ as follows:
\begin{equation}
\left(\frac{d\tau}{d\theta}\right)^{\!2}
 = (1 - f^2) \left(\frac{dt}{d\theta}\right)^{\!2}
   - \frac{1}{c^2}\left[\frac{1}{\lambda^2 (1 - f^2)}
                        \left(\frac{dr}{d\theta}\right)^{\!2} + r^2\right] \, ,
\label{eqn41}
\end{equation}
so
\begin{align}
\left(\frac{dt}{d\theta}\right)^{\!2}
 &= \frac{1}{c^2 (1 - f^2)} \left[c^2 \left(\frac{d\tau}{d\theta}\right)^{\!2}
    + \frac{1}{\lambda^2 (1 - f^2)}
      \left(\frac{dr}{d\theta}\right)^{\!2} + r^2\right] 
\label{eqn42} \\
 &= \frac{\lambda^2}{c^2 [\lambda^2 (1 - f^2)]^2}
    \left\{\lambda^2 (1 - f^2)
           \left[c^2 \left(\frac{d\tau}{d\theta}\right)^{\!2} + r^2\right]
           + \left(\frac{dr}{d\theta}\right)^{\!2}\right\} \, .
\label{eqn43}
\end{align}
After substitutions from $d\tau/d\theta = -r^2/c \, h$ and Eqs.~(\ref{eqn5}),
(\ref{eqn33}), and (\ref{eqn36}) this reduces to
\begin{equation}
\left(\frac{dt}{d\theta}\right)^{\!2}
 = \frac{\lambda^2 a^2 b^2 (h^2 + a^2)(h^2 + b^2)}
        {c^2 h^2 [h^2 + a^2 (\sin\theta)^2 + b^2 (\cos\theta)^2]^2} \, ,
\label{eqn44}
\end{equation}
which yields
\begin{align}
t_{\text B} - t_{\text A}
  = \bigintssss_{\delta}^{-\delta}
    \!\! \frac{dt}{d\theta} \, d\theta
 &= \frac{1}{c \, h} \!\! \bigints_{-\delta}^{\delta}
     \!\!\!\!\!\!\! \frac{\lambda \, a b \sqrt{(h^2 + a^2)(h^2 + b^2)}}
                 {h^2 + a^2 (\sin\theta)^2 + b^2 (\cos\theta)^2} \, d\theta
\label{eqn45} \\
 &= \frac{2 \lambda \, a b}{c \, h}
    \tan^{-1} \! \left(\sqrt{\frac{h^2 + a^2}{h^2 + b^2}}
                       \, \tan(\delta)\right) \, .
\label{eqn46}
\end{align}
The flight time $T_{\epsilon = 1}$ of the particle as read on a clock at B
perfectly synchronized with a matching clock at A is given by
$T_{\epsilon = 1} = \sqrt{1 - f^2(R)} \, (t_{\text B} - t_{\text A})$. 

From the metric $d\sigma^2 = d\rho^2 + r^2(\rho) d\Omega^2$
of $\Sigma_{\bar t}$ and Eq.~(\ref{eqn33}) one gets
\begin{align}
\left(\frac{d\sigma}{d\theta}\right)^{\!2}
 &= \left(\frac{d\rho}{d\theta}\right)^{\!2} + r^2
  = \frac{1}{\lambda^2} \left(\frac{dr}{d\theta}\right)^{\!2} + r^2
\label{eqn47} \\
 &= \frac{1}{\lambda^2} \frac{r^2 (r^2 - a^2)(b^2 - r^2)}
                             {a^2 \, b^2} + r^2
\label{eqn48} \\
 &= \frac{r^6}{\lambda^2 a^2 b^2}
    \left[\left(1 - \frac{a^2}{r^2}\right) \left(\frac{b^2}{r^2} - 1\right)
          + \frac{\lambda^2 a^2 b^2}{r^4}\right] \, .
\label{eqn49}
\end{align}
Substitution from Eq.~(\ref{eqn36}) produces for the length $L_{\epsilon = 1}$
of the particle's path
\begin{equation}
L_{\epsilon = 1}
 = \frac{a b}{\lambda} \! \bigints_{-\delta}^{\delta} \!\!\!\!\!
    \sqrt{\frac{a^4 (\sin\theta)^2
                + \left(\lambda^2 - 1\right)
                  \left[a^2 (\sin\theta)^2 + b^2 (\cos\theta)^2\right]^2
                + b^4 (\cos\theta)^2}
               {\left[a^2 (\sin\theta)^2
                      + b^2 (\cos\theta)^2\right]^3}} \, d\theta \, .
\label{eqn50}
\end{equation} 
The formula for $\bar T_{\epsilon = 1}$ is like that of Eq.~(\ref{eqn25}) for
$\bar T_{\epsilon = 0}$, viz.,
\begin{equation}
\bar T_{\epsilon = 1} \coloneqq \bar t_{\text B} - \bar t_{\text A}
 = t_{\text B} - t_{\text A}
   - \frac{2}{c} \! \bigintss_a^R \!\!\!
                    \frac{\sqrt{\lambda^2 - 1 - (\lambda m/R^3) r^2}}
                         {1 + (\lambda m/R^3) r^2} \, dr \, . 
\label{eqn51}
\end{equation}

\subsection*{Application: Neutrino flight from CERN to Gran Sasso}

To apply these results to the flight from A to B of a particle such as a
neutrino one needs two equations to determine the constants $h$ and $k$.  One
of these will be Eq.~(\ref{eqn37}), the other must involve the particle's rest
mass $m_0$ and the value at A of its energy $E$, related by the well-known
formula $\hat E := E/m_0 c^2 = 1/\sqrt{1 - v^2/c^2}$, where at each event on
the particle's path $v$ is the magnitude of its coordinate three-velocity with
respect to an inertial observer at that event.  At every event such an observer
$\cal O$ is one that is falling freely from rest at $\rho = \infty$ with no
angular motion, whose coordinate four-velocity is
$\partial_{\bar t} + f(\rho) \, c \, \partial_{\rho}$.
The particle's coordinate four-velocity is
$\partial_{\bar t} + (d\rho/d{\bar t}) \, \partial_{\rho}
                   + (d\theta/d{\bar t}) \, \partial_{\theta}$.
Their relative coordinate three-velocity is thus 
$[d\rho/d{\bar t} - f(\rho) \, c] \, \partial_{\rho}
 + (d\theta/d{\bar t}) \, \partial_{\theta}$, the
square of whose magnitude $v$ as measured in the metric of $\Sigma_{\bar t}$
(the metric of space as seen by $\cal O$) is given by
$v^2 = [d\rho/d{\bar t} - f(\rho) \, c \,]^2
       + r^2(\rho) (d\theta/d{\bar t} \, )^2$.
From Eq.~(\ref{eqn4}) one gets
\begin{align}
1 &= \left(\frac{d\bar t}{d\tau}\right)^{\!2}
     - \frac{1}{c^2} \left[\frac{d\rho}{d\tau}
                           - f(\rho) \, c \, \frac{d\bar t}{d\tau}\right]^2
     - \frac{1}{c^2} \, r^2(\rho) \left(\frac{d\theta}{d\tau}\right)^{\!2}
\label{eqn52} \\
  &= \left(\frac{d\bar t}{d\tau}\right)^{\!2}
     \left\{1 - \frac{1}{c^2}
                \left[\frac{d\rho}{d\bar t} - f(\rho) \, c\right]^2
              - \frac{1}{c^2} \, r^2(\rho)
                \left(\frac{d\theta}{d\bar t}\right)^{\!2} \right\}
\label{eqn53} \\
  &= \left(\frac{d\bar t}{d\tau}\right)^{\!2}
     \left(1 - \frac{v^2}{c^2}\right) \, ,
\label{eqn54}
\end{align}
so $\hat E = |d\bar t/d\tau| = |\dot{\bar t}\,|
           = |\dot t - (1/c) f(\rho) [1 - f^2(\rho)]^{-1} \dot \rho|$.
From Eq.~(\ref{eqn52}) follows
$[\dot\rho - f(\rho) \, c \, \dot{\bar t} \, ]^2
 = c^2 \, \dot{\bar t}\,^2 - r^2(\rho) \dot\theta^2 - c^2$,
and then from Eq.~(\ref{eqn7})
\begin{equation}
\dot\rho - f(\rho) \, c \, \dot{\bar t}
  = \sqrt{c^2 \dot{\bar t}\,^2 - r^2(\rho) \dot\theta^2 - c^2}
  = c \, \sqrt{\hat E^2 - \frac{h^2}{r^2(\rho)} - 1} \, ,
\label{eqn55}
\end{equation}
where the positive root is chosen to account for the fact that as time goes on
($\dot{\bar t} > 0$) the particle descends into $\cal B$ more slowly than does
the free-falling observer $\cal O$ ($f(\rho) \, c < d\rho/d{\bar t} < 0$).  Now
Eq.~(\ref{eqn8}) gives
$k = \hat E + f(\rho) \sqrt{\hat E^2 - h^2/r^2(\rho) - 1}$,
which evaluated at~A becomes
\begin{equation}
k = \hat E_0 + f(R) \sqrt{\hat E_0^2 - \frac{h^2}{R^2} - 1} \, ,
\label{eqn56}
\end{equation}
where $\hat E_0 = E_0/m_0 c^2$, the ratio of the initial energy of the particle
to its rest energy.  Solution of Eqs.~(\ref{eqn37}) and~(\ref{eqn56}) for $h$
and $k$ will enable computation of $\tau_{\text B} - \tau_{\text A}$,
$t_{\text B} - t_{\text A}$, $T_{\epsilon = 1}$, $L_{\epsilon = 1}$, and
$\bar T_{\epsilon = 1}$ for various choices of $\hat E_0$.

Squaring both sides of Eq.~(\ref{eqn37}) produces
$\alpha^2 - \beta^2 = R^2 \alpha + R^2 \beta \sin(2 \delta)$
$\left(\text{from } a = \sqrt{\alpha - \beta} \text{ and }
                    b = \sqrt{\alpha + \beta}\,\right)$.
Transposing the term $R^2 \alpha$ and squaring again one arrives ultimately at
\begin{align}
&\left[4 R (R + \lambda m) + \lambda^2 m^2 \sin^2(2 \delta)\right] h^4
\notag \\
&\qquad\quad
 - 2 R^3 \left\{\left(\lambda^2 k^2 - 1\right) \left[2 R + \lambda  m
   \sin^2(2 \delta)\right] - 2 \lambda  m \cos^2(2 \delta)\right\} h^2
 + R^6 \left(\lambda^2 k^2 - 1\right)^2 \sin^2(2 \delta) = 0 \, .
\label{eqn57}
\end{align}
A similar treatment of Eq.~(\ref{eqn56}) produces
\begin{equation}
\left[\lambda^2 R - (R  + \lambda  m)\right] h^2
 + R^2 \left[\lambda^2 R \left(k^2 - 2 {\hat E}_0 k + 1\right)
             + (R + \lambda m)\left({\hat E}_0^2 - 1\right)\right] = 0 \, .
\label{eqn58}
\end{equation}
Substitution of $h^2$ from the second of these equations into the first
produces a polynomial equation of degree four in $k$, so numerical solution is
advised.

In the experiments described in \cite{opera1} and \cite{icarus} neutrinos are
collected at point(s) B (Gran Sasso), having been launched from point A (CERN)
with energy $E_0 \approx 17$ GeV.  The euclidean distance $d = 731$ km
(rounded) from A to B determined from satellite and ground measurements, the
only unknown datum is the neutrino rest energy.  An upper bound on this energy
is thought to be 2 eV.  Taking this for $m_0 c^2$ makes
$\hat E_0 = 8.5 \times 10^9$ (thus the neutrino initial speed
$v_0 = c \, \sqrt{1 - 1/\hat E_0^2} = (1 - 7. \times 10^{-21}) \, c$), for
which choice $k = 8.49998 \times 10^9$ and $h = 5.40642 \times 10^{18}$ cm.
Use of these in the formulas above gives 
\begin{itemize}
\item neutrino's proper time elapsed in flight from A to B:
      \begin{equation}
      \tau_{\text B} - \tau_{\text A} = 2.86866 \times 10^{-13} \text{ sec}
      \label{eqn59}
      \end{equation}
\item neutrino's flight time measured by clock at B synchronized with clock
      at~A:
      \begin{equation}
      T_{\epsilon = 1} = d/c + 9.31085 \times 10^{-16} \text{ sec}
      \quad (= T_{\epsilon = 0} + 1.68745 \times 10^{-23} \text{ sec})
      \label{eqn60}
      \end{equation}
\item neutrino's flight time from A to B as measured by free-falling inertial
      observers:
      \begin{equation}
      \bar T_{\epsilon = 1} = d/c - 2.61131 \times 10^{-9} \text{ sec}
      \quad (= \bar T_{\epsilon = 0} + 1.68745 \times 10^{-23} \text{ sec})
      \label{eqn61}
      \end{equation}
\item length of neutrino's flight path from A to B as measured by
      free-falling inertial observers:
      \begin{equation}
      L_{\epsilon = 1} = d - 8.38501 \times 10^{-5} \text{ cm}
                       = d - 0.838501 \; \mu\text{m}
      \quad (= L_{\epsilon = 0} + 1.61152 \times 10^{-33} \text{ cm})
      \label{eqn62}
      \end{equation}
\item average speed of neutrino in flight from A to B as measured by
      free-falling inertial observers:
      \begin{equation}
      \frac{L_{\epsilon = 1}}{\bar T_{\epsilon = 1}} 
      = c + 0.32106 \text{ km/sec}
      = 1.00000107093 \, c = (1 + 1.07093 \times 10^{-6}) \, c
      \quad \left(= \frac{L_{\epsilon = 0}}{\bar T_{\epsilon = 0}}
                    - 2.07470 \times 10^{-10} \text{ cm/sec}\right)
      \label{eqn63}
      \end{equation}
\end{itemize}
That $T_{\epsilon = 1}$ exceeds $d/c$ by $9.31085 \times 10^{-16} \text{ sec}$
is the result of relevance to the experiments described in \cite{opera1}
and~\cite{icarus}.

\subsection*{Application: Newton's cannonball} 

Isaac Newton imagined a cannon firing a cannonball horizontally from a high
mountaintop with velocity just sufficient to keep it from ever falling to
ground.  If we bring his cannon down to a point A on the ball $\cal B$ (taken
to represent a nonrotating, homogeneous, spherical Earth along whose surface
the cannonball can travel without hindrance), then the formulas derived above
will apply with $a = b = R$, in which case we have that $r(\rho) = \rho = R$,
\begin{itemize}
\item that $\alpha - \beta = a^2 = R^2$ and $\alpha + \beta = b^2 = R^2$, thus
      $\alpha = R^2$ and $0 = \beta
      = \sqrt{\alpha^2 - R^3 h^2/\lambda m}
      = \sqrt{R^4 - R^3 h^2/\lambda m}$, and therefore
      $h = \sqrt{\lambda m R} = 1.68093 \times 10^4$ cm,
\item from Eqs.~(\ref{eqn7}), (\ref{eqn8}), and (\ref{eqn9}) that
      $k^2 = (1 - f^2(R))(1 + h^2/R^2)$, thus that
      $k = (1/\lambda)(1 + \lambda m/R) = 0.9999999997
      = 1 - 3. \times 10^{-10}$, and
\item from $k = \hat E + f(R) \sqrt{\hat E^2 - h^2/R^2 - 1}$ and
      Eq.~(\ref{eqn56}) that $\hat E = \hat E_0 = \lambda = 1.000000001
      = 1 + 1. \times 10^{-9}$.
\end{itemize}
These give, as measured by free-falling inertial observers, a flight time for
the cannonball's `round' trip from A to A of
$84.34771$ min $= 1$ hr $24$ min $20.86261$ sec, a flight path length
of $4.00302 \times 10^4$ km $= 2 \pi R$, and an average (in fact, a constant) 
speed of $7.90975$ km/sec.  For a 12 lb cannonball of inertial rest mass
$m_0 = 12 \text{ lb}/g$ in common use in Newton's day the amount of gunpowder
required to send it on its way with the required kinetic energy
(= total energy - rest energy = $E_0 - m_0 c^2 = (\hat E_0 - 1) m_0 c^2
= 3.18861 \times 10^{18}$~GeV = $3.76758 \times 10^8$ ft-lb) would be about
377 pounds (at 500 ft-tons per pound of powder~\cite{noble}).

\subsection{Spacelike geodesics}

For a spacelike geodesic parametrized by the arclength parameter
$\hat\tau \coloneqq i \tau$, $\epsilon = -1$ and Eq.~(\ref{eqn12}) reduces to
\begin{equation}
\dot\rho^2
 = \frac{c^2 m}{\lambda R^3 r^2}
    \left(r^4 + 2 \, \bar\alpha \, r^2 - \frac{R^3 h^2}{\lambda m}\right) \, ,
\label{eqn64}
\end{equation}
where
\begin{equation}
\bar\alpha = \frac{(\lambda^2 k^2 + 1) R^3 - \lambda m h^2}{2 \lambda m} \, .
\label{eqn65}
\end{equation}
If $h = 0$, then $\dot\theta = 0$ and the geodesic follows a diameter of
$\cal B$ from one end to the other.  If $h \neq 0$, then Eq.~(\ref{eqn64})
combined with Eq.~(\ref{eqn7}) produces
\begin{equation}
\left(\frac{dr}{d\theta}\right)^{\!2}
 = \lambda^2 \frac{\dot\rho^2}{\dot\theta^2}
 = \frac{\lambda m}{R^3 h^2} \, r^2 (r^2 - \bar a^2)(r^2 + \bar b^2)
 = \frac{r^2 (r^2 - \bar a^2)(r^2 + \bar b^2)}{\bar a^2 \, \bar b^2} \, ,
\label{eqn66}
\end{equation}
where $\bar a = \sqrt{\bar\beta - \bar\alpha}$,
      $\bar b = \sqrt{\bar\beta + \bar\alpha}$, and
      $\bar\beta = \sqrt{\bar\alpha^2 + R^3 h^2/\lambda m}$.
This implies that $0 < \bar a \leq r$, and if $r = \bar a$ when $\theta = 0$,
then, as in Eq.~(\ref{eqn34}),
\begin{align}
\theta
 &= \sgn(\theta) \!\! \bigintssss_{\bar a}^r \!\! \frac{d\theta}{ds} \, ds
  = \sgn(\theta) \!\!\! \bigintss_{\bar a}^r \!\!\!\!\!
                 \frac{\bar a \, \bar b}
                      {s \sqrt{(s^2 - \bar a^2)(s^2 + \bar b^2)}} \, ds
\label{eqn67} \\
 &= \sgn(\theta)
    \tan^{-1} \! \left(\frac{\bar b}{\bar a}
                       \sqrt{\frac{r^2 - \bar a^2}{r^2 + \bar b^2}}\right) \, ,
\label{eqn68}
\end{align}
from which follows
\begin{equation}
r = \frac{\bar a \, \bar b}
         {\sqrt{\bar b^2 \, (\cos\theta)^2 - \bar a^2 \, (\sin\theta)^2}} \, .
\label{eqn69}
\end{equation}
This describes a trajectory inside $\cal B$ that in euclidean geometry, where
$x = r \cos\theta$ and $y = r \sin\theta$, would be an an arc of a hyperbola
centered on the center point $\cal C$ of $\cal B$, with transverse ($x$) axis
of length $2 \bar a$ and conjugate ($y$) axis of length $2 \bar b$.  The
initial conditions $r = R$ and $\theta = \delta$ at A produce from
Eq.~(\ref{eqn69}) 
\begin{equation}
\bar a \, \bar b
 = R {\sqrt{\bar b^2 \, (\cos\delta)^2 -
            \bar a^2 \, (\sin\delta)^2}} \, ,
\label{eqn70}
\end{equation}
which yields the constraint $\bar b/\bar a > \tan(\delta)$ on the axes of the
hyperbola.  Consequent on this equation and
$\bar a^2 \bar b^2 = \bar\beta^2 - \bar\alpha^2 = R^3 h^2/\lambda m$ is
\begin{equation}
h^2 = \frac{\lambda m}{R^3} \bar a^4 \frac{R^2 - r_0^2}{r_0^2 - \bar a^2} \, ,
\label{eqn71}
\end{equation}
from which follow that $\bar a < r_0$ and $h \to \infty$ as $\bar a \to r_0$
(and vice versa).

From $d\theta/d\hat\tau = \dot\theta = -c \, h/r^2$ and Eq.~(\ref{eqn69}) one
can find the proper time elapsed on the particle's clock (if in fact there is
a particle following the geodesic, and the particle has a clock, and $\hat\tau$
is the time measured by that clock) as follows:
\begin{align}
\hat\tau_{\text B} - \hat\tau_{\text A}
  = \int_{\hat\tau_{\text A}}^{\hat\tau_{\text B}} \!\!\! d\hat\tau
  = -\frac{1}{c \, h} \int_{\delta}^{-\delta} \!\!\! r^2 \,  d\theta
 &= \frac{\bar a^2 \bar b^2}{c \, h} \!
    \bigintssss_{-\delta}^{\delta}
       \frac{1}{\bar b^2 \, (\cos\theta)^2
                - \bar a^2 \, (\sin\theta)^2} \, d\theta
\label{eqn72} \\
 &= \frac{2 \, \bar a \bar b}{c \, h}
    \left[\tanh^{-1} \!
    \left(\frac{\bar a}{\bar b} \tan(\delta)\right)\right] \, .
\label{eqn73}
\end{align}
To find the proper time elapsed on a clock at B synchronized with a clock at A
requires computing $t_{\text B} - t_{\text A}$.  A calculation like that
leading up to Eq.~(\ref{eqn44}) shows that
\begin{equation}
\left(\frac{dt}{d\theta}\right)^{\!2}
 = \frac{\lambda^2 \bar a^2 \bar b^2 (h^2 - \bar a^2)(h^2 + \bar b^2)}
        {c^2 h^2 [h^2 - \bar a^2 (\sin\theta)^2
                      + \bar b^2 (\cos\theta)^2]^2} \, ,
\label{eqn74}
\end{equation}
from which follows that $h \geq \bar a$ and
\begin{equation}
t_{\text B} - t_{\text A}
 = \pm\frac{2 \lambda \, \bar a \bar b}{c \, h}
      \tan^{-1} \! \left(\sqrt{\frac{h^2 - \bar a^2}{h^2 + \bar b^2}}
                         \tan(\delta)\right) \, .
\label{eqn75}
\end{equation}
The geodesic with $h = \bar a$, on which no time $t$ passes in the particle's
trip from A to B, separates the geodesics on which the particle arrives before
it started from those on which it arrives after it started.  The flight time
$T_{\epsilon = -1}$ of the particle as read on a clock at B perfectly
synchronized with a matching clock at A is given by
$T_{\epsilon = -1} = \sqrt{1 - f^2(R)} \, (t_{\text B} - t_{\text A})$. 

A calculation like that for Eq.~(\ref{eqn50}) produces for the length
$L_{\epsilon = -1}$ of the particle's path
\begin{equation}
L_{\epsilon = -1}
 = \frac{\bar a \bar b}{\lambda} \!
   \bigints_{-\delta}^{\delta} \!\!\!
      \sqrt{\frac{\bar a^4 (\sin\theta)^2
                  + \left(\lambda^2 - 1\right)
                    \left[\bar b^2 (\cos\theta)^2 -
                          \bar a^2 (\sin\theta)^2\right]^2
                  + \bar b^4 (\cos\theta)^2}
                 {\left[\bar b^2 (\cos\theta)^2
                        - \bar a^2 (\sin\theta)^2\right]^3}} \, d\theta \, .
\label{eqn76}
\end{equation} 
For the inertially measured flight time of the particle from A to B the analog
of Eqs.~(\ref{eqn25}) and (\ref{eqn51}) is
\begin{equation}
\bar T_{\epsilon = -1} \coloneqq \bar t_{\text B} - \bar t_{\text A}
 = t_{\text B} - t_{\text A}
   - \frac{2}{c} \! \bigintss_{\bar a}^R \!\!\!
                    \frac{\sqrt{\lambda^2 - 1 - (\lambda m/R^3) r^2}}
                         {1 + (\lambda m/R^3) r^2} \, dr \, . 
\label{eqn77}
\end{equation}

\section{Discussion}
\enlargethispage{0.5in}

Suppose a photon $\gamma$ and a neutrino $\nu$ depart from a point A at time 0
and arrive at a point B at times $T_\gamma$ and $T_\nu$ as measured by a clock
at B perfectly synchronized with a matching clock at A.  If $T_\nu < T_\gamma$,
is one entitled to say that the neutrino traveled faster than the photon?  One
is not, for missing is any information about the lengths $L_\gamma$ and $L_\nu$ 
of the paths that the particles followed.  In the interpretation of the results
of the experiment described in~\cite{opera1} it was assumed (in the absence of
other, more realistic options) that $L_\gamma = L_\nu = d$, the euclidean
distance from the neutrino source point A at CERN's European Laboratory for
Particle Physics to a point B in the OPERA detector at LNGS (Laboratori
Nazionale del Gran Sasso), and that $T_\gamma = d/c$.  $T_\nu$ was reported to
have been observed to be less than $T_\gamma$ by about
57.8 ns.\footnote{Corrected at a later date to
$T_\nu \approx T_\gamma - 6.5$ ns \cite{opera2} and thereby brought into
approximate compatibility with the results reported in \cite{icarus}.}  This
was interpreted to imply that the neutrino's speed exceeded that of light by
about $(2.4 \times 10^{-5}) \, c \approx 7.2$ km/sec.  On its face this is not
an allowable inference, as it compares the speed of the neutrino traveling
through the gravitational field inside Earth to the speed of a photon traveling
through empty space.  Given, however, that the actual distance a photon would
travel through a tunnel between A and B would likely differ very little from
$d$, and that its speed in the tunnel should differ very little from $c$, the
inference was not unreasonable.


A proper comparison between $T_\gamma$ and $T_\nu$ must have the photon and the
neutrino travel in the same space-time geometry, such as that inside the Earth
as depicted in this paper.  Even then the comparison cannot be exact, as the
particles follow different routes from A to B, but in the applications detailed
above for the CERN to Gran Sasso measurements the maximum separation between
their routes is $(a - r_0)/\lambda = 1.00849 \times 10^{-23}$ cm and the
neutrino's route is only $1.61152 \times 10^{-33}$ cm longer than the photon's, so the comparison is nearly exact.  The model predicted that the neutrino's
flight time would exceed the photon's by $1.68745 \times 10^{-23}$ cm/sec,
whether measured by the clocks stationed at A and B or by clocks carried by
free-falling inertial observers (allowed to penetrate Earth without interacting
nongravitationally with its matter).  As seen in Eqs.~(\ref{eqn26}),
(\ref{eqn27}), (\ref{eqn60}), and (\ref{eqn61}), these times would be greater
than $T_\gamma$ (the time a photon would take traveling in a vacuum) by about
$10^{-15}$ sec as measured by the surface clocks, but less than $T_\gamma$ by
about $10^{-9}$ sec as measured by the inertial clocks.  Moreover, as seen in
Eq.~(\ref{eqn63}), the average speeds of the neutrino and the photon as
measured in the inertial frames along their paths, differing from one another
by about $10^{-10}$ cm/sec, exceed $c$ by about 321 meter/sec.  This somewhat
unintuitive result merely reflects the fact that the geometry inside Earth
differs in a particular way from the geometry in a vacuum.  The way it differs
is determined here by the new, improved field equations employed to discover
and govern it \cite{elli1}, field equations created to correct Einstein's
unjustified assumption that inertial-passive mass (and therefore energy) can
masquerade as active gravitational mass in the production of
gravity~\cite{elli2}.

Modeling Earth as a nonrotating, homogeneous, spherical ball in order to
analyze photon and neutrino flights from CERN to Gran Sasso is, of course,
dictated by the relative ease of solving the field equations~(\ref{eqn1}) under
those restrictions.  Let us consider the possibility of relaxing those
restrictions and what the effects of doing so might be.
\begin{itemize}
\item Allowing inhomogeneity while retaining the other restrictions could be 
      accomplished by numerically solving the field equations with a radially
      varying density $\mu$ such as that profiled in~\cite{besi} (based
      on~\cite{dzhala}).  Subsequent numerical computations of the various
      integrals in Secs.~I and II would likely yield for $d = 731$ km results
      differing very little from those found here, inasmuch as the maximum
      depth of the photon's trajectory is $(R - r_0)/\lambda =
      (R - R \cos(\delta))/\lambda = 10.5$ km, and that of the neutrino's
      trajectory is less.  For neutrino detectors contemplated as targets more
      distant from CERN, such as Majorana Demonstrator in South Dakota and
      Super-Kamiokande in Japan, the results might be significantly different
      from those for a constant density $\mu$. 
\item To take into account Earth's rotation one would ideally find an interior
      solution of the field equations that matched up at the surface with some
      solution of the vacuum field equations that could reasonably be
      interpreted as modeling the gravitational field exterior to a rotating
      Earth.  This would likely require giving up spherical symmetry in favor
      of an oblate spheroidal symmetry, which Earth has to a close
      approximation.  If the exterior solution were taken to be a portion of
      the Kerr space-time manifold~\cite{kerr}, finding a matching interior
      solution might be feasible with the density $\mu$ constant.  Otherwise
      the problem would reduce to numerically solving partial differential
      equations in two variables, $\rho$ and $\vartheta$ (or $\theta$).  The
      corrections to flight times of photons and neutrinos would likely be
      relatively small.
\end{itemize}
There are other variations to be taken into account, most notably the
elevations above sea level of the starting and ending points of the photon and
neutrino trajectories, and the mountains and valleys above the flight paths.
These have been examined in detail in~\cite{besi}.
\enlargethispage{0.35in}

\begin{center} \rule{3.5in}{.01in} \end{center}

\noindent
Homepage: \url{http://euclid.colorado.edu/~ellis}
\hfill
Email: {ellis@euclid.colorado.edu}

\end{document}